\documentstyle[aps,prl,preprint]{revtex}
\newcommand{\be}{\begin{equation}}
\newcommand{\ee}{\end{equation}}
\newcommand{\bea}{\begin{eqnarray}}
\newcommand{\eea}{\end{eqnarray}}

\newcommand{\nonum}{\nonumber}
\tightenlines
\author{J.~U.~N{\"o}ckel and K.~Richter}
\address{Max-Planck-Institut f{\"u}r Physik komplexer Systeme\\
N{\"o}thnitzer Str. 38, 
01187 Dresden, Germany}
\title{AC transport with reservoirs of finite width}
\begin{document}
\maketitle
\date{\today}
\draft
\date{\today}
\begin{abstract}
The linear response conductance coefficients are calculated 
in the scattering approach at finite
frequency, damping and magnetic field for a microstructure in which the
reservoirs are modeled as quantum wire leads of infinite length but finite
width. Independently of frequency, inelastic scattering causes 
subbands with large group velocity to 
contribute more strongly to the conductance
than channels of comparable transmission but slower propagation. At finite
frequency and magnetic fields, additional correction terms appear, some of 
which are sensitive to the phase of the S matrix.
\end{abstract}

The Landauer-B{\"u}ttiker formula \cite{lbformprl} for the low-temperature, 
linear-response  DC conductance coefficients of a 
small conductor requires only the knowledge of the quantum mechanical 
transmission coefficients, i.e. absolute values of S-matrix elements. 
The S-matrix in turn appears because one necessarily has to consider the
system as being open, i.e. having a continuous spectrum. With discrete 
levels, no dissipative response is possible.  One often uses a 
model in which the sample is connected to the outside world by 
straight quantum wires of infinte length \cite{baranger}. 

Here, we want to study the linear conductance of such an open system under 
an AC perturbation.
The limiting case of infinitely broad reservoirs and no damping 
has been considered previously, 
\cite{buttikerprl93}, yielding an {\em admittance matrix} for $n\ne l$ 
\be\label{buttikereqn}
g_{nl}(\omega)=\frac{e^{2}}{h}\,\int\!\!dE\,\frac{f(E)-f(E+\hbar\omega)} 
{\hbar\omega}
\,{\rm Tr}\left\{S^{\dagger}_{nl}(E)\,S_{nl}(E+\hbar\omega)\right\}.
\ee
Here, $n$ and $l$ label the leads so that 
$\langle I_{n}(\omega)\rangle=\sum\limits_{l}\,g_{nl}(\omega)\,V_{l}
(\omega)$, and the trace extends over subbands.
An apparent conflit has been noted \cite{buttikerchristen95} between 
the widespread use of Landauer-type formulas even in systems that are 
one-dimensional at infinity, and the requirement that the reservoirs 
have to be much wider than the sample itself to permit equilibration.

We provide an answer to this open question by considering the linear
response of a system with quantum wire leads, including an inelastic
scattering rate in the equation of motion. Electron-electron interactions are
assumed to be absent in the reservoir region. An
external uniform magnetic field $B$ is taken into account. 

Consider a system 
under the influence of an external perturbation, $H = H_0 + H_1$, 
where $H_0$ is the unperturbed time-independent Hamiltonian and (in 
the interaction picture)
\be\label{scalargaugeeqn}
H_{1I}(t)\equiv -e\int n_{I}({\bf x},t)\,\Phi({\bf x})\,d{\bf x}\equiv
e^{i\omega t}\,F_I(t).
\ee
Here, $n_I(t)$ is the number density operator. 
The assumption that a static potential $\Phi$ can describe the field 
distribution amounts to neglecting the rotational part of the 
time-varying electric field. 
The fact that the resulting perturbation (which must also contain the 
self-consistent induced fields) just adds 
to the original Hamiltonian $H_0$ is a consequence 
of the linear response approximation. 
The expectation value of the current density is
$\langle j({\bf x},\,t)\rangle = Tr\,\rho(t)\,j$,
where we use the {\em linear approximation} in the von-Neumann 
equation for the density matrix, 
\begin{equation}\label{lin1}
\frac{\partial}{\partial t}\rho_I(t) = \frac{1}{i\hbar}\,\left[
H_{1I}(t),\rho_0\right] - \gamma\,\left\{\rho_I(t)-\rho_0\right\}.
\end{equation}
Here $\rho_0$ is the equilibrium density and $\gamma$ is an {\em 
inelastic relaxation rate} \cite{trivedi,yariv}. To solve this in the 
basis of many-body eigenstates $\vert\alpha\rangle$ of $H_0$ with 
energy $\epsilon_{\alpha}$, one needs the matrix elements of 
Eq.\ (\ref{scalargaugeeqn}). Using the continuity equation, 
${\dot n}_{I}({\bf x},t)=-\nabla\cdot{\bf j}_{I}({\bf x},t)$, one can 
perform an integration by parts 
as in Ref.\ \cite{noeckelbr93}, yielding
\bea
\langle\alpha\vert F_{I}(t)\vert\beta\rangle
&=&\frac{ie\hbar}{\epsilon_{\alpha}-\epsilon_{\beta}}
\,\left\{\int\limits_{\partial{\cal A}}
\langle\alpha\vert{\bf j}_{I}({\bf x},t)\vert\beta\rangle\,
\Phi({\bf x})\,d{\bf x}\right.\nonum\\
&&\left.-\int\limits_{\cal A}
\Phi({\bf x})\,\nabla\cdot\langle\alpha\vert
{\bf j}_{I}({\bf x},t)\vert\beta\rangle\,d{\bf x}\right\}.
\label{startperteqn}
\eea
Here, ${\bf j}_I$ is the current operator associated with $H_0$, 
containing at most a static vector potential if the Landau gauge 
is chosen in each lead\cite{baranger}. 

With this procedure, we have traded the infinite integration domain 
of Eq.\ (\ref{scalargaugeeqn}) for a finite domain $\cal A$.
The price that was paid is the additional boundary integral over 
$\partial{\cal A}$. 
At $\omega=\gamma=0$, the contribution of the bulk term is known to vanish
\cite{baranger}. Its 
significance at finite frequency as compared to the surface 
term is not addressed in Ref.\ \cite{buttikerprl93}, but it has been 
stated\cite{janssen} that it should be negligible at small 
frequencies, when one is in the quasi-stationary regime, $\nabla {\bf 
j}\approx0$. However, an estimate of the conditions for this regime is 
not available in the literature.

Using $\langle\alpha\vert\rho\vert\beta\rangle$, the current response becomes
\be\label{javereqn2}
\langle{\bf j}({\bf x},\,t)\rangle={\bf j}_0({\bf x}) - e^{i\omega t}\,
\sum\limits_{\alpha\beta}
\frac{
\langle\beta\vert{\bf j}_I({\bf x},\,t)\vert\alpha\rangle\,
\langle\alpha\vert F_I(t)\vert\beta\rangle
}{\hbar\omega+\epsilon_{\alpha}-\epsilon_{\beta}+i\hbar\gamma}
\,(P_{\alpha}-P_{\beta}).
\ee
Here, ${\bf j}_0({\bf x})$ is the unperturbed equilibrium current, 
and $P_{\alpha}=\langle\alpha\vert\rho_0\vert\alpha\rangle$. 
The total current is the integral of Eq.\ (\ref{javereqn2}) 
over the cross section of lead $n$. 
Furthermore, we only need the deviation from the equilibrium value, 
so that we bring the term involving ${\bf j}_0({\bf x})$ to the 
lefthand side. The resulting induced current has the time dependence 
$e^{i\omega t}$, which we drop to get the corresponding Fourier 
component (which can still be a 
function of the distance $x$ from the sample). 

For the bulk term in Eq.\ (\ref{startperteqn}), we can calculate the
total current through all the leads,
\bea
\sum\limits_n I_n^{\rm bulk}&=&\int\limits_{\cal A}\nabla\cdot
\langle{\bf j}^{\rm bulk}({\bf x'})-{\bf j}_0^{\rm bulk}({\bf x'})\rangle
d{\bf x}'
\nonum\\
&=&-\frac{ie^2}{\hbar}\sum\limits_{\alpha\beta}\frac{P_{\beta}-P_{\alpha}}
{\epsilon_{\beta}-\epsilon_{\alpha}-\hbar
z}(\epsilon_{\beta}-\epsilon_{\beta})\times\nonum\\
&&\int\limits_{\cal A}\langle\beta\vert 
n({\bf x'})\vert\alpha\rangle\,\Phi({\bf x})\,
\langle\alpha\vert n({\bf x})\vert\beta\rangle\,d{\bf x}'\,d{\bf x},
\eea
where Gauss' theorem and the continuity equation were used. We henceforth
abbreviate $z\equiv\omega- i\gamma$. To linear order in $z$, this can be
expressed in terms of the Green's function, 
$G(E)=\sum_{\alpha}\vert\alpha\rangle\langle\alpha\vert/
(E-\epsilon_{\alpha})$,
as 
\be
\sum\limits_n I_n^{\rm bulk}=-ie^2z\,
\sum\limits_{\beta}P_{\beta}
\int\limits_{\cal A}\Phi({\bf x})\,\langle\beta\vert
n({\bf x}')\,G(\epsilon_{\beta})\,n({\bf x})+{\rm h.c.\,}\vert\beta\rangle
\,d{\bf x}'\,d{\bf x}.
\ee
The similarity of this operator expression to the mean-field-like result of
Ref.\ \cite{buttikerchristen95} [cf.\ Eq.\ (54) therein, summed over leads]
leads us to identify this contribution with the internal response, which we
have thus re-derived in an alternative way. 

In the following we shall focus on the boundary term 
and find a generalization of the external response result of 
Ref.\ \cite{buttikerchristen95,remark}. 
The $N$ mutually decoupled, straight
infinite quantum wires leads at large distance from the sample 
are described by a non-interacting Hamiltonian $H_{QW}$. 
The many-body eigenstates $\vert\alpha\rangle$ 
of the full unperturbed system decompose into Slater determinants of 
single-particle eigenstates of $H_{QW}$ in the
asymptotic region, a complete set of which is given by the
scattering states, $\psi^l_{Eap}$ at energy $E$, consisting of an incoming
wave in subband $a$ of lead $p$, and of outgoing waves in all the subbands of
all the leads $l$. Using local coordinates $x$, $y$ parallel and 
transverse to the wire $l$, the energy-normalized quantum wire 
eigenfunctions are
\begin{equation}\label{efn}
\xi_{\pm a}^l({\bf x})\equiv\left|2\pi\,\frac{dE_a^l}{dk}\right|^{-1/2}
e^{i{\rm k}_{\pm a}^lx}\,\chi_{{ \rm k}_{\pm a}^la}^l(y).
\end{equation}
Unless $B=0$ in the leads, the transverse 
wavefunctions $\chi^{l}_{ka}(y)$ are also functions of the wavenumber 
${\rm k} _{\pm a}^l (E)$. The $+$ $(-)$ sign denotes propagation toward (away
from) the sample. The scattering states in lead $l$ then are
\begin{equation}\label{scatt}
\psi_{Ea{p}}^l\equiv
\delta_{{p} l}
\,\xi_{-a}^l({\bf x})+\sum_{a'}S_{l{p},a'a}\,\xi_{+a'}^l.
\end{equation}
Here, $S_{l{p},a'a}$ is the $S$ matrix element for scattering from
subband $a$ of lead $p$ to subband $a'$ of lead $l$. 
Bound states of the system do not enter in the
completeness relation because of their exponential decay. 

After making the transition to the single-particle scattering states, the
boundary contribution to the current at a distance $x$ along lead $n$
from the sample takes the form
\bea
I_n(x)&=&
-ie^2\hbar\,\sum\limits_{l=1}^N V_l
\sum\limits_{p,q=1}^N
\sum\limits_{a,b=0}^{\infty}\,\int\limits_{-\infty}^{\infty}\!\!dE'
\,dE
\,\frac{f(E')-f(E)}{E-E'}\nonum\\
&&
\times\frac{
\langle\psi_{Ea{p}}^l\vert J^l(x_{\cal A})\vert\psi_{E'b{q}}^{l}\rangle
\,
\langle\psi_{E'b{q}}^{n}\vert J^n(x)\vert\psi_{Ea{p}}^n
\rangle}{E'-E-\hbar z}.\label{singleparteqn1}
\eea
Here, $J^l$ is the component of the single-particle current operator 
along lead $l$, integrated
over the cross section at a distance $x_{\cal A}$ from the sample (determined
by $\partial{\cal A}$). We use the convention that subscripts $a$ or
$a'$ are always associated with a dependence on $E$, while $b$ or $b'$ 
labels functions of $E'$. 
If we now insert Eq.\ (\ref{scatt}), then the last line in 
Eq.\ (\ref{singleparteqn1}) contains a product of current 
matrix elements between quantum wire eigenfunctions at different energies, of
the type considered in Ref.\ \cite{noeckelbr93}. For brevity, we restrict
ourselves to $g_{nl}$ with $n\ne l$. 
In that case one can exploit the $\delta_{pl}$ in 
Eq.\ (\ref{scatt}), as well as the unitarity relation 
$S^{\dagger}(E)\,S(E)=1$. 
Furthermore, current matrix elements between counter-propagating waves such as
$\langle\xi_{-a}^l\vert J^l\vert\xi_{+b}^l\rangle$
vanish as $\omega\to 0$. To linear order in $\omega$, this leaves
\bea
g_{nl}&=&
-ie^2\hbar\,\sum\limits_{a,b,a',b'=0}^{\infty}\,
\int\limits_{-\infty}^{\infty}\!\!dE'\,dE
\,\frac{f(E')-f(E)}{E-E'}\nonum\\
&&
\times\frac{1}{E'-E-\hbar z}\,\left\{
\langle\xi_{-a}^l\vert J^l\vert\xi_{-b}^{l}\rangle\,
\langle\xi_{+b'}^n\vert J^n\vert\xi_{+a'}^{n}\rangle
\,S^*_{nl,b'b}\,S_{nl,a'a}\right.\nonum\\
&&+\left.
\langle\xi_{+a'}^l\vert J^l\vert\xi_{+b'}^{l}\rangle\,
\langle\xi_{-b}^n\vert J^n\vert\xi_{-a}^{n}\rangle\,
S^*_{ln,a'a}\,S_{ln,b'b}\right\}
\label{singleparteqn2}
\eea
The arguments on the current operators were left out, 
keeping in mind that the coordinates 
are $x_{\cal A}$ in leads $l$ and $x$ in lead $n$. 
Following Ref.\ \cite{buttikerprl93}, one of the energy integrals
is done such that all terms multiplied by $f(E)$ are integrated 
over $E'$, {\em vice versa} for terms multiplied by $f(E')$. Applying the 
residue theorem, the poles of the Fermi function then do not  
enter. If the first term in braces is integrated 
over $E$ ($E'$), we close the contour by a large semicircle in 
the positive (negative) imaginary plane; for the second term in 
braces, the opposite contour is taken. With this choice, all the 
$e^{\pm ikx}$ 
factors in the wire eigenfunctions $\xi$ yield exponential 
suppression along the return contour due to the imaginary part of 
$k$. Poles of $S$ occur only in the negative 
energy plane, and in the positive plane for $S^*$. As a consequence, 
the above choice of contours always selects the half plane in 
which no poles of the S-matrix are enclosed. The contributions from 
the pole at $E'=E$ cancel due to the difference of Fermi functions, 
so that the only pole that remains is the one at $E'=E+\hbar z$. 
However, this pole is never enclosed for the second term in braces, so 
that the latter makes no contribution at all. We are thus left with
\bea
g_{nl}&=&
2\pi\,e^2\hbar\,\sum\limits_{a,b,a',b'=0}^{\infty}\,
\int\limits_{-\infty}^{\infty}\!\!dE
\,\frac{f(E+\hbar z)-f(E)}{\hbar z}\nonum\\
&&
\times\left.
\langle\xi_{-a}^l\vert J^l\vert\xi_{-b}^{l}\rangle\,
\langle\xi_{+b'}^n\vert J^n\vert\xi_{+a'}^{n}\rangle
\,S^*_{nl,b'b}\,S_{nl,a'a}\right|_{E'=E+\hbar z}.
\label{condcoeffeqn}
\eea
At
$B=0$, the transverse modes are orthogonal irrespective of energy, 
yielding immediately the requirement $a=b$, $a'=b'$.
Consider these terms first, but at $B\ne 0$. 
Expand all wavenumbers
around $E$, recalling that $b$ labels functions of $E+\hbar z$, e.g.
\be
{\rm k}_{-b}^l(E+\hbar z)\approx{\rm k}_{-b}^l(E)+q^l_{-a},\quad
q^l_{-a}\equiv\frac{z}{v_{g,-a}^l(E)}.
\label{groupvelexpeqn2}
\ee
Here, we introduced the channel-specific group velocity 
$
v_{g,\pm a}^l(E)\equiv(1/\hbar)\,
dE_{a}^l({\rm k}_{\pm a}^l(E))/dk$. 
Then first-order perturbation theory yields for the transverse channel 
eigenfunctions 
\be
\chi^l_{-b,E+\hbar z}=\chi^l_{-a}-\hbar\omega_c\,q^l_{-a}\,\sum\limits_{a'\ne a}
\chi^l_{-a'}\frac{\langle\chi^l_{-a'}\vert y'\vert\chi^l_{-a}\rangle}
{E^l_{-a}({\rm k}^l_{-a}(E))-E^l_{-a'}({\rm k}^l_{-a}(E))},
\ee
where $\omega_c=eB/mc$. This can be used to evaluate the 
current matrix elements in Eq.\ (\ref{condcoeffeqn}) with the Landau gauge, 
where one needs
$\langle\chi^l_{-a}\vert\chi^l_{-b}\rangle$, 
$\langle\chi^l_{-a}\vert y'\vert\chi^l_{-b}\rangle$ and 
$\langle\chi^n_{+b'}\vert\chi^n_{+a'}\rangle$, 
$\langle\chi^n_{+b'}\vert y\vert\chi^n_{+a'}\rangle$. 
To get explicit expressions, we specialize to {\em
parabolic} quantum wires with the same dispersion relation in leads $n$ and
$l$, 
\be
E_a(k)=\frac{\hbar^2k^2}{2m}\,\frac{\omega_0^2}{\omega_0^2+\omega_c^2}+
\hbar\,\sqrt{\omega_0^2+\omega_c^2}\,\left(a+\frac{1}{2}\right).
\ee
The terms in Eq.\ (\ref{condcoeffeqn}) with
$a=b$, $a'=b'$ then yield to linear order in $z$
\bea
g_{nl}^{diag}&=&
\frac{e^2}{h}\,\sum\limits_{a,a'=0}^{\infty}\,
\int\limits_{-\infty}^{\infty}\!\!dE
\,\left[-\frac{f(E+\hbar z)-f(E)}{\hbar z}\right]\,
S_{nl,a'a}^*(E+\hbar z)\,S_{nl,a'a}(E)
\label{condcoeffdiageqn}\\
&&\times e^{-i\,z\,\left(
x/|v^n_{ga'}|+x_{\cal A}/|v^l_{ga}|\right)}\,
\left\{1+z\,\frac{m\omega_c^2}{2\hbar\omega_0^2}\,
\left(\frac{1}{(k^n_{a'})^2}-\frac{1}{(k^l_{a})^2}\right)\right.\nonum\\
&&\left.\qquad\qquad\qquad\qquad\qquad\times
\left(1+\frac{\omega_c^2}{\omega_0^2}\right)
\,\left(1+\sqrt{1 + \omega_c^2/\omega_0^2}\right)\,
\right\}\nonum
\eea
This reduces to Eq.\ (\ref{buttikereqn}) only if {\em both} $\omega_c=0$
{\em and} one drops the exponential, which describes damped oscillations 
as a function of $x$ and $x_{\cal A}$ since $z=\omega-i\,\gamma$. 
Note that $x_{\cal A}\,\gamma$ is the 
velocity that a particle must have in order to traverse the distance 
$x_{\cal A}$ before the inelastic damping becomes significant. Similarly, 
$x_{\cal A}\,\omega$ is the velocity that is required to 
traverse $x_{\cal A}$ in one oscillation period of the external field. 
If carriers can then enter and leave the sample region so fast that
$z\,x/|v^n_{ga'}|$, $z\,x_{\cal A}/|v^l_{ga}\approx 0$, then the 
precise location of the  current and voltage probes, which is given 
by $x$ and $x_{\cal A}$, respectively, becomes irrelevant. In 
particular, channels exceeding the {\em cutoff} group velocity, 
max$(x\gamma,\,x_{\cal A}\gamma)$, are essentially undamped in 
Eq.\ (\ref{condcoeffdiageqn}). Their number increases when the 
reservoirs are much wider than the sample region, in which case ``slow'' 
subbands typically have negligible transmission through the sample 
since coupling to the narrow sample requires large transverse momentum 
transfer. For narrow reservoirs, the transmission of such slow 
subbands can be appreciable, so that the deviation between 
Eqs.\ (\ref{buttikereqn}) and (\ref{condcoeffdiageqn}) can be 
significant even at $\omega=B=0$. 

We also find contributions from {\em off-diagonal terms} with $a\ne b$ 
or $a'\ne b'$ in Eq.\ (\ref{condcoeffeqn}), of which we list only one 
example\cite{future}:
\bea
g_{nl}^{off}&=&
z\,\frac{e^2}{h}\,
\frac{\omega_c}{\omega_0^2}\,
\sqrt{\omega_0^2+\omega_c^2}\,\sqrt{\frac{\hbar}{2\,m\,\omega_0}}\,
\frac{m}{\hbar}\,
\sum\limits_{a,a'=0}^{\infty}\,
\int\limits_{-\infty}^{\infty}\!\!dE
\,\left[-\frac{f(E+\hbar z)-f(E)}{\hbar z}\right]
\times\nonum\\
&&\frac{\sqrt{a+1}}
{\sqrt{{\rm k}_{-(a+1)}^l\,{\rm k}_{-a}^l}}\,
e^{-i\,z\,\left(
x/|v^n_{ga'}|+x_{\cal A}/|v^l_{ga+1}|\right)}\,
e^{i\,\left({\rm k}_{-(a+1)}^l-{\rm k}_{-a}^l\right)\,x_{\cal A}}\,
S_{nl;a',a+1}^*(E+\hbar z)\,S_{nl,a'a}(E)
\nonum\\
\eea
This contains further oscillatory exponentials, which for 
large $x_{\cal A}$ lead to suppression of the integral. 
However, this may be compensated by the fact that 
$g_{nl}^{off}$ has roughly one power of wavenumber more than 
$g_{nl}^{diag}$. In that case the admittance at $B\ne 0$ can give 
information about products of S-matrix elements that cannot be written 
as a trace over subbands as in Eq.\ (\ref{buttikereqn}).

We benefited from discussions with 
M.~B{\"u}ttiker, M.~Leadbeater, E.~McCann and A.~D.~Stone.


\end{document}